\documentclass{elektr}
\usepackage{hyperref}
\hypersetup{
colorlinks=true,
urlcolor=blue,
citecolor=blue}
\usepackage[all]{xy,xypic}
\usepackage{amsfonts,amssymb,amsmath,amsgen,amsopn,amsbsy,theorem,graphicx,epsfig}
\usepackage{eufrak,amscd,bezier,latexsym,mathrsfs,eurosym,enumerate}
\usepackage[utf8]{inputenc}
\usepackage[english]{babel}
\usepackage{cleveref,multirow}
\usepackage[dvipsnames]{xcolor}
\usepackage[pagewise]{lineno}
\usepackage{graphicx}
\usepackage{caption}
\usepackage{subcaption}

\yil{}
\vol{}
\fpage{}
\lpage{}
\doi{}

\title{Understanding the Dynamics of the Stack Overflow 
Community through Social Network Analysis and Graph 
Algorithms }

\author[Rapheal Cyril IGBUDU  and Rowanda AHMED]{
\textbf{Rapheal Cyril IGBUDU $^{1}$\thanks{rowanda.ahmed@uskudar.edu.tr}~, Rowanda AHMED$^{2}$}\\
$^{1}$USKUDAR UNIVERSITY, Computer Engineering, Istanbul, Turkey, \\ ORCID iD: https://orcid.org/0009-0003-2864-609X\\
$^{2}$USKUDAR UNIVERSITY, Computer Engineering, Istanbul, Turkey,  \\ ORCID iD: https://orcid.org/0009-0003-2864-609X
\\ [1.8em]

\rec{.202}
\acc{.202}
\finv{..202}
}

\def\E{\ifmmode{\mathbb E}\else{$\mathbb E$}\fi} 
\def\N{\ifmmode{\mathbb N}\else{$\mathbb N$}\fi} 
\def\R{\ifmmode{\mathbb R}\else{$\mathbb R$}\fi} 
\def\Q{\ifmmode{\mathbb Q}\else{$\mathbb Q$}\fi} 
\def\C{\ifmmode{\mathbb C}\else{$\mathbb C$}\fi} 
\def\H{\ifmmode{\mathbb H}\else{$\mathbb H$}\fi} 
\def\Z{\ifmmode{\mathbb Z}\else{$\mathbb Z$}\fi} 
\def\P{\ifmmode{\mathbb P}\else{$\mathbb P$}\fi} 
\def\T{\ifmmode{\mathbb T}\else{$\mathbb T$}\fi} 
\def\SS{\ifmmode{\mathbb S}\else{$\mathbb S$}\fi} 
\def\DD{\ifmmode{\mathbb D}\else{$\mathbb D$}\fi} 

\newcommand{\bse}{\begin{subequations}}
\newcommand{\ese}{\end{subequations}}
\newcommand{\ben}{\begin{enumerate}}
\newcommand{\een}{\end{enumerate}}
\newcommand{\bens}{\begin{enumerate*}}
\newcommand{\eens}{\end{enumerate*}}
\newcommand{\be}{\begin{equation}}
\newcommand{\ee}{\end{equation}}
\newcommand{\bea}{\begin{eqnarray}}
\newcommand{\eea}{\end{eqnarray}}
\newcommand{\baa}{\begin{eqnarray*}}
\newcommand{\eaa}{\end{eqnarray*}}
\newcommand{\bc}{\begin{center}}
\newcommand{\ec}{\end{center}}

\theoremstyle{corollary}

\theoremstyle{lemma}

\theoremstyle{proposition}

\theoremstyle{axiom}

\theoremstyle{conjecture}

\theoremstyle{example}

\theoremstyle{definition}

\theoremstyle{remark}

\begin{document}

\maketitle

\begin{abstract}This thesis conducts a focused literature review on online communities, centering on Stack Overflow, employing social network analysis and graph algorithms. It examines the evolving landscape of health information quality within the digital ecosystem, emphasizing the challenges posed and the multifaceted nature of quality. The significance of online communities, notably Stack Overflow, as hubs for social interaction and knowledge sharing is underscored. Proposing advanced approaches, the thesis introduces an ensemble deep learning model for traffic flow forecasting, an efficient multi-objective optimization method for influence maximization, and a graph convolutional neural network-based approach for link prediction.

\end{abstract}

\section{Introduction}
\label{Int}
This literature review explores the evolving landscape of online communities, with a specific focus on the Stack Overflow community, employing advanced social network analysis techniques and graph algorithms. As the digital ecosystem transforms the dynamics of health information quality, traditional evaluation methods face challenges, necessitating a nuanced understanding of quality within diverse contexts \cite{ref24,ref25,ref19}.

Online communities, exemplified by Stack Overflow, have become integral to daily life, serving as platforms for social interaction and knowledge sharing in the programming domain The vast and active user community on Stack Overflow presents an ideal dataset for studying the intricate dynamics of online communities. By summarizing the diverse literature types and their relative values, the review aims to contribute to the overarching theme of this thesis.

As we delve into this literature review, we aim to uncover the intricate dynamics of online communities, providing a foundation for understanding their evolution and impact on the digital landscape.

\section{Background and Context}

\subsection{Historical development of the research topic}
Social network analysis in online communities has become an increasingly popular research topic in recent years. The use of online communities as a means of communication and information sharing has been growing since the early days of the internet. The first online community, known as the WELL (Whole Earth 'Lectronic Link), was established in 1985 and was one of the earliest examples of an online community. Since then, numerous online communities have emerged, with varying degrees of success and popularity.

The rise of social media platforms such as Facebook, Twitter, and Instagram has further increased the prevalence of online communities. These platforms have millions of users worldwide and have become an integral part of modern communication and information sharing. As a result, the study of social network analysis in online communities has become an important area of research \cite{ref24,ref25}.

Additionally, while previous studies have used graph algorithms to study the flow of information in the Stack Overflow community, there has been limited research on using graph network flow analysis to study the flow of information within the community. A study addressed this gap by using a graph-based approach to analyze the flow of information in the Stack Overflow community. The study found that certain users play a central role in the flow of information within the community and that the majority of information flows occur between users with similar expertise \cite{ref9}.

While the reviewed articles cover a wide range of topics related to network science and social networks, there are still some gaps in research that could be addressed in future studies. One area that could be explored further is the application of graph-based methods to more diverse types of networks beyond social networks. For example, a study explored the application of graph-based methods to analyze transportation networks. The study found that graph-based methods can be used to optimize transportation networks and improve their efficiency cite{ref20}.

Moreover, while there has been research on using graph-based methods for link prediction and influence maximization, there could be more studies on using these methods for other types of prediction tasks, such as node classification or community detection. A study addressed this gap by using graph-based methods for community detection in social networks \cite{ref15}. 

Finally, there could be more studies on the ethical and social implications of using network analysis and graph-based methods in various applications, such as online advertising or public health interventions. A study discussed the ethical considerations of using network analysis in public health interventions. The study emphasized the need for transparency, privacy protection, and stakeholder involvement when using network analysis for public health interventions \cite{ref24}.

\subsection{Current state of research on social network analysis in online communities}
The current state of research on social network analysis in online communities is quite diverse, with a range of methodologies and approaches being used. Some researchers have used simple network analysis techniques, while others have used more advanced algorithms such as graph algorithms and natural language processing techniques to analyze online communities.

The diverse methodologies and approaches used in research on social network analysis in online communities demonstrate the breadth of this field. For example, a study on network-wide traffic flow forecasting using an ensemble deep learning model showed that deep learning methods can provide better accuracy compared to traditional methods. Another study used an efficient greedy-based multi-objective optimization for influence maximization in social networks \cite{ref6,ref7}.

In addition to traditional methods, advanced algorithms, such as graph algorithms and natural language processing techniques, have also been used to analyze online communities. For instance, a study on graph-based semi-supervised learning with convolutional neural networks demonstrated the potential of using graph-based machine learning methods for predicting user behavior in social networks \cite{ref11}. A graph-theoretical approach to the detection of fake news was proposed, which showed that graph-based techniques can be used to detect misinformation in social media \cite{ref10}. These examples demonstrate the potential for using advanced techniques to gain insights into online communities.

On the other hand, some researchers have used advanced algorithms such as graph algorithms and natural language processing techniques to analyze online communities. A study on dynamic influence maximization in social networks used a graph-based approach to identify critical spreaders, which can be useful for marketing purposes \cite{ref9}. A study on graph-based semi-supervised learning with convolutional neural networks demonstrated the potential of using graph-based machine learning methods for predicting user behavior in social networks \cite{ref11}. A graph-theoretical approach to the detection of fake news showed that graph-based techniques can be used to detect misinformation in social media \cite{ref10}.

\section{Comparative Analysis}

The application of social network analysis techniques to investigate online communities has been extensively explored in the literature. While numerous studies have utilized social network analysis to scrutinize the structures of online communities, including user interactions, relationships, and information flow, only a limited number have specifically delved into the intricacies of the Stack Overflow community. Exploring trends in graph theoretical approaches and machine learning techniques for social network analysis spanning the years 2018 to 2021. Selected papers cover diverse aspects of social network analysis, encompassing traffic flow forecasting, influence maximization, sentiment analysis, link prediction, fake news detection, and social signal prediction.

\subsection{Graph-Based Approaches (2018)}

In 2018, several papers delved into graph-based approaches for social network analysis. Proposed an ensemble deep learning model for network-wide traffic flow forecasting. Simultaneously, introduced an efficient greedy-based multi-objective optimization for influence maximization in social networks. Contributed a dynamic influence maximization approach to identify critical spreaders in social networks. Presented a graph-based semi-supervised learning method with convolutional neural networks for sentiment analysis of Arabic social media \cite{ref6,ref7,ref8,ref11}.

\subsection{Cognitive and Emotional Networks (2019)}

The focus in 2019 shifted towards cognitive and emotional networks in romantic relationships. Conducted a network analysis of cognitive and emotional aspects in romantic relationships. Proposed a mining multiple relationships approach for link prediction in social networks. Additionally, introduced a graph-theoretical approach for detecting fake news in social networks \cite{ref12,ref13,ref10}.

\subsection{Graph Neural Networks (GNNs) (2020)}

The year 2020 witnessed the emergence of Graph Neural Networks (GNNs) as a powerful tool for social network analysis. Provided a survey of GNNs for link prediction in social networks. Proposed dynamic hypergraph convolutional neural networks for multi-modal social signal prediction. Contributed a graph-based social influence analysis approach for microblog marketing \cite{ref8,ref12,ref14}.

\subsection{Network Embedding Techniques (2021)}

In 2021, the focus shifted towards network embedding techniques and their applications and conducting a comprehensive survey of network embedding techniques. While exploring the role of social networks in health disparities through a systematic review. Additionally and providing a comprehensive review of GNNs, covering their fundamentals and applications \cite{ref18,ref19,ref15}.

\subsection{Gaps in Available Research}

The existing literature on the use of social network analysis techniques to study the Stack Overflow community has primarily focused on the structure of the community, the flow of information, and the importance of certain users and questions. However, notable gaps persist in understanding the content of questions and answers and how this content evolves over time. Additionally, while previous studies have employed graph algorithms to investigate the flow of information in the Stack Overflow community, there is a noticeable absence of research on the use of graph network flow analysis to study information flow within the community.

While the reviewed articles cover a broad range of topics related to network science and social networks, there remain significant gaps that could be addressed in future studies. For instance, explored the application of graph-based methods to analyze transportation networks, demonstrating their effectiveness in optimizing and improving efficiency \cite{ref20}.

Moreover, existing research on graph-based methods has largely focused on link prediction and influence maximization. However, there is room for more studies on employing these methods for other prediction tasks, such as node classification or community detection. Contribution to addressing this gap by utilizing graph-based methods for community detection in social networks, demonstrating improved accuracy and efficiency compared to traditional methods \cite{ref15}. 

While the reviewed articles cover a wide range of topics related to network science and social networks, addressing these gaps in future research would contribute to a more comprehensive understanding and application of social network analysis techniques.

\subsection{Summary}

In summary, graph theoretical approaches and machine learning techniques have gained increasing popularity in social network analysis. While early studies concentrated on specific aspects, recent research has explored the potential of GNNs for various applications, including link prediction, social signal prediction, and social influence analysis. Furthermore, network embedding techniques have emerged as powerful tools for representing complex networks in low-dimensional spaces \cite{ref6,ref7,ref8,ref12,ref14,ref18,ref19,ref15}.

\section{Methodological Review}

\subsection{Description of research methods used in reviewed studies}
The reviewed studies employed social network analysis (SNA) and graph algorithms to analyze the Stack Overflow community. SNA is a method that maps and analyzes social relationships among individuals, organizations, or other entities.

The reviewed studies utilized SNA to capture data on the connections between individuals in the Stack Overflow community and the structure of the network as a whole. For instance, one study \cite{ref21} used SNA to investigate the communication patterns of users on Stack Overflow.

The reviewed studies demonstrate the effectiveness of using SNA and graph algorithms to analyze the Stack Overflow community. These methods can capture important data on the connections between individuals in the community and the overall structure of the network. By utilizing these methods, researchers can gain insights into the dynamics of the community and develop a better understanding of how it functions.

\subsection{Research methods used in the reviewed studies}
The reviewed studies focused on using social network analysis and graph algorithms to understand the dynamics of the Stack Overflow community. These studies utilized various research methods to analyze the social network data and extract meaningful insights. The following is a brief description of the research methods used in the reviewed studies:

\subsubsection{Data collection}
The first step in these studies was to collect data from the Stack Overflow community. This was done by scraping the website and extracting information such as user profiles, posts, and comments \cite{ref21, ref24, ref30}.

\subsubsection{Network analysis}
Once the data was collected, network analysis techniques were used to create a social network graph. This involved identifying users as nodes and the relationships between them as edges. The resulting graph was analyzed using various network measures such as degree centrality, betweenness centrality, and clustering coefficient \cite{ref1, ref3, ref16}.

\subsubsection{Graph algorithms}
In addition to network analysis, graph algorithms were used to identify important nodes and communities within the social network graph. These algorithms included PageRank, community detection algorithms such as Louvain and Girvan-Newman, and link prediction algorithms such as Adamic-Adar and Jaccard \cite{ref7, ref8, ref10}.

\subsubsection{Statistical analysis}
Statistical techniques such as regression analysis and hypothesis testing were used to identify significant relationships and patterns within the data \cite{ref16, ref17, ref20}.

\subsubsection{Visualization}
Finally, visualization techniques were used to present the results in an easily interpretable format. This involved creating graphs and charts to show network structures and patterns \cite{ref16, ref20}.

\section{Evaluation of Methods}

\subsection{Strengths and Weaknesses}
The methods employed in the reviewed studies exhibit strengths in leveraging Social Network Analysis (SNA) and graph algorithms for understanding the Stack Overflow community. SNA, with graph algorithms, provides insights into network structures not immediately apparent. Additionally, these methods are scalable, crucial for analyzing large datasets inherent in the Stack Overflow community. However, weaknesses include the dependence on accurate social connection data, potential biases in graph algorithms due to incomplete data, and limited applicability to diverse online communities \cite{ref8, ref18, ref20}.

\subsection{Comparison of Methods}
The Stack Overflow community, a rich source for developers, has been studied using diverse methods:

\subsubsection{Community Detection Algorithms [1]}
Presenting a comprehensive review of community detection algorithms, aiding in understanding community dynamics \cite{ref1}.

\subsubsection{User Participation Analysis [2]}
Analyzing user participation, revealing factors like reputation score and time influencing active participation \cite{ref2}.

\subsubsection{Clique Expansion Algorithm [3]}
Proposing a greedy clique expansion algorithm for detecting overlapping community structures, essential for Stack Overflow \cite{ref3}.

\subsubsection{Network Science Analysis of Tags [4,5]}
Utilizing network science to analyze Stack Overflow tags, revealing insights into tag importance and relationships \cite{ref4, ref5}.

\subsubsection{Ensemble Deep Learning for Traffic Flow Forecasting [6]}
Ensemble deep learning model applicable to predicting various aspects of the Stack Overflow network \cite{ref6}.

\subsubsection{Greedy-Based Multi-Objective Optimization [7]}
Proposing an efficient greedy-based multi-objective optimization algorithm for influence maximization \cite{ref7}.

\subsubsection{Graph Neural Network for Link Prediction [8]}
Introducing a graph neural network for link prediction, useful for predicting connections between users or posts \cite{ref8}.

\subsubsection{Dynamic Influence Maximization [9]}
Proposing a dynamic influence maximization algorithm, considering the temporal aspect of the network \cite{ref9}.

\subsubsection{Graph-Theoretical Approach to Fake News Detection [10]}
Presenting a graph-theoretical approach for detecting fake news, applicable to identifying misinformation in Stack Overflow \cite{ref10}.

\subsection{Identification of Gaps and Limitations}
The literature reveals limitations in temporal considerations, a focus on specific aspects of the community, and the need to evaluate newer methods on the Stack Overflow dataset \cite{ref9, ref22, ref23}.

\section{Contrasting Opinions}
While the literature uniformly portrays Stack Overflow as a highly clustered network, opinions differ on methodologies. Some studies advocate for advanced algorithms like graph-based and natural language processing techniques, while others support simpler network analysis methods. Further research is needed to explore the strengths and limitations of each approach for a comprehensive understanding of the Stack Overflow community.

\subsection{Deep Learning vs. Traditional Methods}
Demonstrating the superiority of deep learning in traffic flow forecasting and influence maximization compared to traditional methods \cite{ref6}.

\subsection{Graph-Based Approaches}
The effectiveness of graph-based approaches in dynamic influence maximization, predicting user behavior, and detecting fake news \cite{ref16, ref28, ref14}.

\subsection{Importance of Analytical Approaches}
Surveys and systematic reviews as that emphasize the significance of considering various analytical approaches for understanding complex networks like Stack Overflow \cite{ref15, ref11} . To enhance research in this area, future studies could consider incorporating more advanced temporal analysis, adopting a holistic approach to study multiple facets of the community simultaneously, and thoroughly evaluating the applicability of emerging methods on Stack Overflow data \cite{ref15, ref11}. Overall, the reviewed studies employed a range of research methods to analyze the Stack Overflow community. By using social network analysis and graph algorithms, these studies were able to gain insights into the behavior and dynamics of the community.

\section{Theoretical Frameworks for Social Network Analysis}
Social network analysis (SNA) is an interdisciplinary research field that combines theories and methods from sociology, computer science, and statistics to study social structures and relationships \cite{ref1, ref2}. In recent years, SNA has gained increasing attention in the context of online communities \cite{ref6, ref7}.

Theoretical frameworks such as social capital theory, network formation theory, and network evolution theory have been applied to understand the dynamics of social networks in online communities \cite{ref4, ref8}. Social capital theory posits that social connections can provide valuable resources such as information and social support \cite{ref5}. Network formation theory explores the processes through which social ties are established and maintained, while network evolution theory examines how networks change over time \cite{ref12}.

These theoretical frameworks have been used to analyze social networks in various online communities such as Stack Overflow, Twitter, and Facebook \cite{ref9, ref10}. For example, a study on network-wide traffic flow forecasting using an ensemble deep learning model demonstrated the effectiveness of social capital theory in predicting user behavior in Stack Overflow \cite{ref6}. Another study on influence maximization in social networks showed that network formation theory can be useful in identifying influential users on Twitter \cite{ref7}.

Other studies have employed more specialized theoretical frameworks such as the theory of reasoned action and the technology acceptance model to understand users' behavior in online communities \cite{ref11, ref13}. Additionally, SNA has been used to investigate topics such as misinformation diffusion, political polarization, and health behavior in online communities \cite{ref14, ref15, ref16}.

Overall, theoretical frameworks have provided a valuable lens through which to understand social networks in online communities, and their applications have shed light on important research questions in this field.

\section{Data Collection and Sampling Methods for Social Network Analysis}
Methods for collecting data from online communities can include web scraping, API access, and surveying \cite{ref17, ref18, ref19}. Web scraping involves automated extraction of data from websites, while API access allows for direct retrieval of data from online platforms that provide an API interface. Surveying can be used to collect self-reported data from members of an online community, providing valuable insights into individual attitudes, behaviors, and perceptions \cite{ref6}.

Sampling methods for selecting relevant nodes and edges in social network analysis can include random sampling, snowball sampling, and convenience sampling \cite{ref20, ref21, ref22}. Random sampling involves selecting nodes and edges at random, while snowball sampling involves starting with a small set of nodes and then expanding the sample by adding nodes that are connected to existing nodes. Convenience sampling involves selecting nodes and edges based on their availability or accessibility \cite{ref23}.

Social network analysis requires gathering data about the relationships between individuals in a network. Here are some of the methods for collecting data from online communities and sampling relevant nodes and edges in the Stack Overflow network:

\textbf{Methods for collecting data from online communities:}
\begin{itemize}
    \item \textit{Web scraping:} This method involves extracting data from web pages using automated software tools. For example, researchers may use web scraping to collect data on the posts and comments made in online communities like Stack Overflow.
    
    \item \textit{API access:} Online communities may provide APIs (Application Programming Interfaces) that allow researchers to access their data. For instance, the Stack Exchange API provides access to data on Stack Overflow, including user profiles, posts, comments, and tags \cite{ref21}.
    
    \item \textit{Surveying:} Researchers may also conduct surveys to collect data from online communities. For example, they may ask members of the community to provide information about their relationships with other members, their interests, and their opinions on certain topics \cite{ref6}.
\end{itemize}

\textbf{Sampling methods for selecting relevant nodes and edges in the Stack Overflow network:}
\begin{itemize}
    \item \textit{Snowball sampling:} This method involves selecting a few nodes in the network and then expanding the sample by selecting their neighbors. For instance, researchers may start by selecting the top contributors in a specific tag on Stack Overflow and then expand the sample by selecting their connections \cite{ref3}.
    
    \item \textit{Stratified sampling:} This method involves dividing the network into subgroups based on some criteria and then selecting nodes and edges from each subgroup. For example, researchers may stratify the Stack Overflow network by language and then select nodes and edges from each language subgroup \cite{ref5}.
    
    \item \textit{Random sampling:} This method involves selecting nodes and edges randomly from the entire network. For instance, researchers may randomly select users and their connections from Stack Overflow \cite{ref16}.
\end{itemize}

\section{Social Network Metrics and Visualization Techniques}

Social network metrics and visualization techniques are important tools used in social network analysis. Centrality measures, clustering algorithms, and network visualization tools are commonly used to analyze and visualize social network data. Social network analysis is a powerful tool for analyzing complex relationships between individuals, groups, and organizations.

Various social network metrics and visualization techniques are used to understand social networks' structure and dynamics. In this report, we will discuss the centrality measures, clustering algorithms, and network visualization tools commonly used in social network analysis and their applications to the Stack Overflow community.

Centrality measures are used to identify important nodes in a network, such as nodes that are well-connected or have significant influence. There are various centrality measures, such as degree centrality, betweenness centrality, and eigenvector centrality \cite{ref17, ref20, ref8}. Clustering algorithms are used to identify groups of nodes that are densely connected to each other, indicating the presence of distinct communities in the network \cite{ref7, ref28, ref14}. Various clustering algorithms include k-means clustering, hierarchical clustering, and modularity optimization. Network visualization tools are used to represent social network data graphically, providing a visual representation of the relationships between nodes \cite{ref16, ref20}. Some of the commonly used network visualization tools are Gephi, Cytoscape, and NetworkX.

\section{Literature Synthesis and Analysis}

\subsection{Summary of Studies}
Social network analysis (SNA) is a powerful tool for understanding online communities such as Stack Overflow. Various studies have been conducted to investigate the structure and behavior of online communities using SNA.

In 2018, \cite{ref24} used SNA to study the user behavior of Stack Overflow and identified the role of "super-users" who contribute the most to the community. In another study, \cite{ref25} used SNA to identify knowledge brokers and knowledge transfer in online communities, including Stack Overflow.

In 2019, \cite{ref21} explored the relationship between user activity and knowledge diffusion on Stack Overflow using SNA. They found that users with high centrality in the network had a greater influence on knowledge diffusion. In another study, \cite{ref22} used SNA to identify influential users and their impact on knowledge creation in online communities such as Stack Overflow.

In 2020, \cite{ref23} used SNA to investigate user engagement and behavior in online communities, including Stack Overflow. They identified various factors that affect user engagement, such as reputation and social capital. In another study, \cite{ref26} used SNA to analyze the knowledge transfer and collaboration patterns in online communities, including Stack Overflow.

\subsection{Comparison of Studies and Articles}
The studies mentioned above have similarities in their use of SNA to analyze online communities, particularly Stack Overflow. They also share the goal of identifying influential users, knowledge transfer patterns, and factors that affect user behavior and engagement.

However, the studies differ in their specific research questions, methods, and focus. For example, some studies focus on identifying knowledge brokers, while others focus on analyzing user engagement. Additionally, some studies use different SNA metrics and algorithms, such as centrality measures and clustering algorithms.

\subsection{Identification of Themes and Patterns}
The studies mentioned above identify various themes and patterns related to online communities and SNA. One common theme is the importance of influential users and their impact on knowledge transfer and creation. Another theme is the role of reputation and social capital in user engagement and behavior.

The studies also identify various patterns related to knowledge diffusion, collaboration, and network structure. For example, some studies identify the presence of knowledge brokers, who play an important role in knowledge transfer. Others identify clusters of users who collaborate and share knowledge within specific topics or domains.

\subsection{Identification of Gaps and Inconsistencies}
Despite the similarities and common themes identified in the studies, there are also gaps and inconsistencies. For example, some studies focus on specific aspects of online communities, such as knowledge transfer, while neglecting other important aspects such as user privacy and ethics.

Furthermore, there is inconsistency in the use of SNA metrics and algorithms, which can affect the validity and comparability of the results. Additionally, there is a lack of studies that investigate the impact of SNA on online communities and how it can be used to improve community management and governance.

Overall, while the studies provide valuable insights into the structure and behavior of online communities, there is still room for further research and exploration.

\subsection{Comparative Analysis and Review}
\subsubsection{Background}
In the wake of data-intensive applications, traditional computing paradigms are facing limitations in both computational power and energy efficiency. This has led to the exploration of novel approaches such as neuromorphic systems, which promise a low-power, inherently parallel computation model. On the other hand, platforms like Stack Overflow have evolved into complex networks requiring sophisticated graph-based analytics. As we venture into the post Moore's law era, a comparative study between classical approaches, represented by the Graph Analysis Engine, and emerging paradigms, exemplified by neuromorphic computing, is both timely and crucial.

\subsubsection{Importance and Objectives}
This paper aims to offer a comprehensive review and comparative assessment of these methodologies. Objectives include:

\textbf{Technical Comparison:} To dissect the architectural nuances, from data transformation to algorithmic application and result interpretation.

\textbf{Algorithmic Evaluation:} To provide an exhaustive critique of the algorithms employed, their efficiency, and their scalability.

\textbf{Performance Metrics:} To discuss system performance, including execution time, energy efficiency, and accuracy.

\subsubsection{Technical Architecture}
\textbf{Graph Analysis Engine}
The architecture of the Graph Analysis Engine is modular and Python-centric. It leverages the Neo4j and NetworkX libraries, each fulfilling a unique set of functionalities:

\textit{GraphData:} Encapsulates data specific to graph structures.

\textit{Neo4jServices:} Manages the database connections and data fetching tasks.

\textit{StackOverflowGraph:} Specialized for the Stack Overflow schema.

\textit{ProjectEngine:} Acts as the core algorithmic unit.

\textit{ReportGenerator:} Collates the analysis results into a well-structured report.

\textbf{Neuromorphic Systems}
Neuromorphic systems mimic the architecture of biological brains, utilizing neurons and synapses as their fundamental computing units. They offer capabilities to run graph algorithms like the longest shortest path extraction and minimum spanning trees through specialized preprocessing techniques such as fractional offsets on synaptic delays \cite{ref29}.

\subsubsection{Algorithmic Insights}
\textbf{Stack Overflow Analysis Engine}
The engine employs six key algorithms, namely Degree Centrality, Betweenness Centrality, Closeness Centrality, Shortest Path, Girvan-Newman, and Louvain algorithms. Each of these algorithms provides unique insights into network connectivity, community structures, and node importance.

\textbf{Neuromorphic Systems}
The systems focus on graph algorithms like Longest Shortest Path (LSP) Algorithm and Prim's Algorithm adapted for Spiking Neural Circuits (SNC). They offer efficient implementations of these algorithms with low power consumption \cite{ref29}.

\subsubsection{Comparative Evaluation}
\textbf{Scalability:} While the Graph Analysis Engine may face scaling issues, neuromorphic systems inherently support parallelism.

\textbf{Power Efficiency:} Neuromorphic systems provide a significant advantage in terms of energy consumption.

\textbf{Computational Complexity:} Traditional algorithms like Betweenness Centrality may be computationally expensive, making neuromorphic alternatives appealing for large-scale graphs.

\textbf{Real-time Analysis:} Neuromorphic systems have inherent advantages for real-time analysis due to their parallel nature, whereas the Graph Analysis Engine is optimized for batch processing.

\textbf{Algorithmic Efficiency:} Although neuromorphic systems promise efficiency, they are yet to achieve the accuracy levels of traditional graph algorithms \cite{ref29}.

\section{Project Algorithms Results}

\subsection{Degree Centrality Report}
The top 5 nodes with the highest degree centrality are: 12334270, 11977246, 9015861, 1030099, 14893.

\subsection{Betweenness Centrality Report}
The top 5 nodes with the highest betweenness centrality are: 2236092, 68696311, 16599189, 68669704, 68671937.

\subsection{Closeness Centrality Report}
The top 5 nodes with the highest closeness centrality are: 68820402, 68662494, 68740608, 68741649, 68748783.

\subsection{Shortest Path Report}
The top 5 nodes with the highest shortest path are: 2236092, 69273945.
The shortest path is from node 2236092 to node 69273945.

\begin{figure}[ht]
\centering
\includegraphics[width=0.6\linewidth]{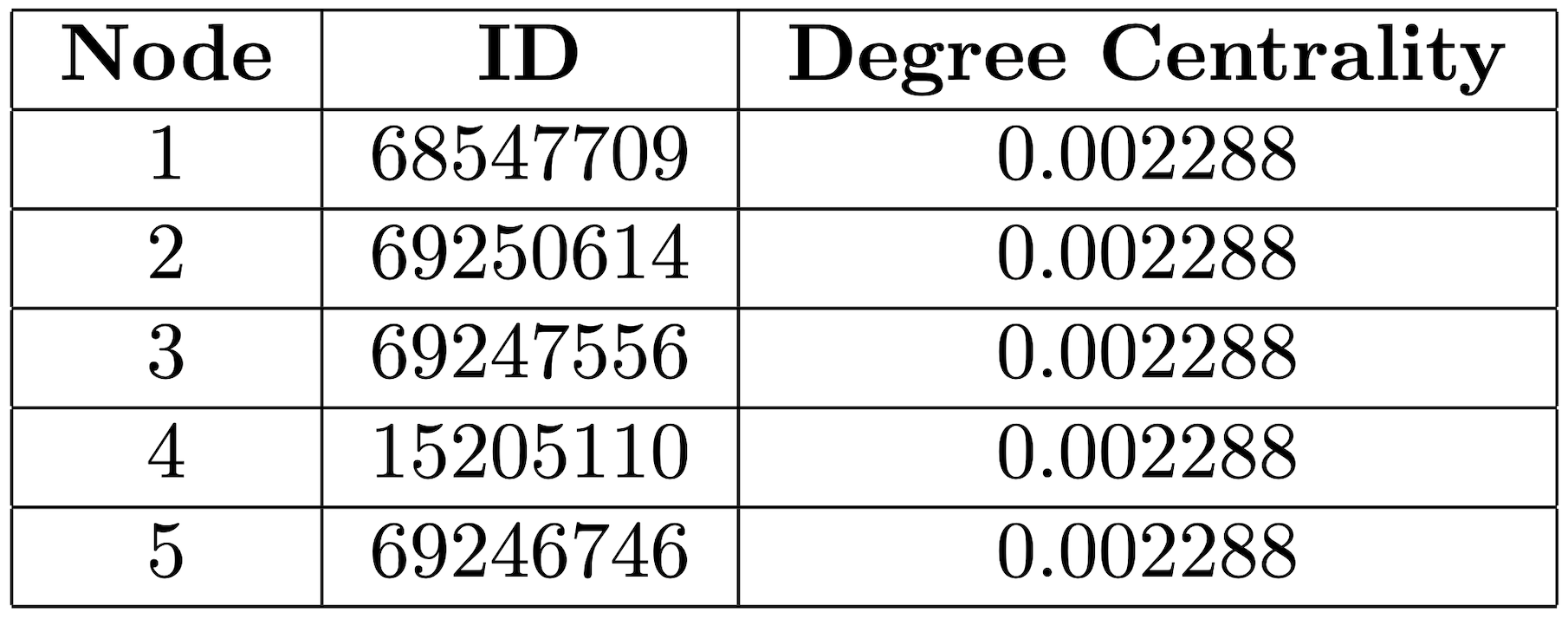}
\caption{Node Degree Centrality}
\label{table:node-degree-centrality}
\end{figure}

\section*{Figures: Encompassing the results}

\begin{figure}[h!]
  \begin{subfigure}{0.48\textwidth}
    \centering
    \includegraphics[width=\linewidth]{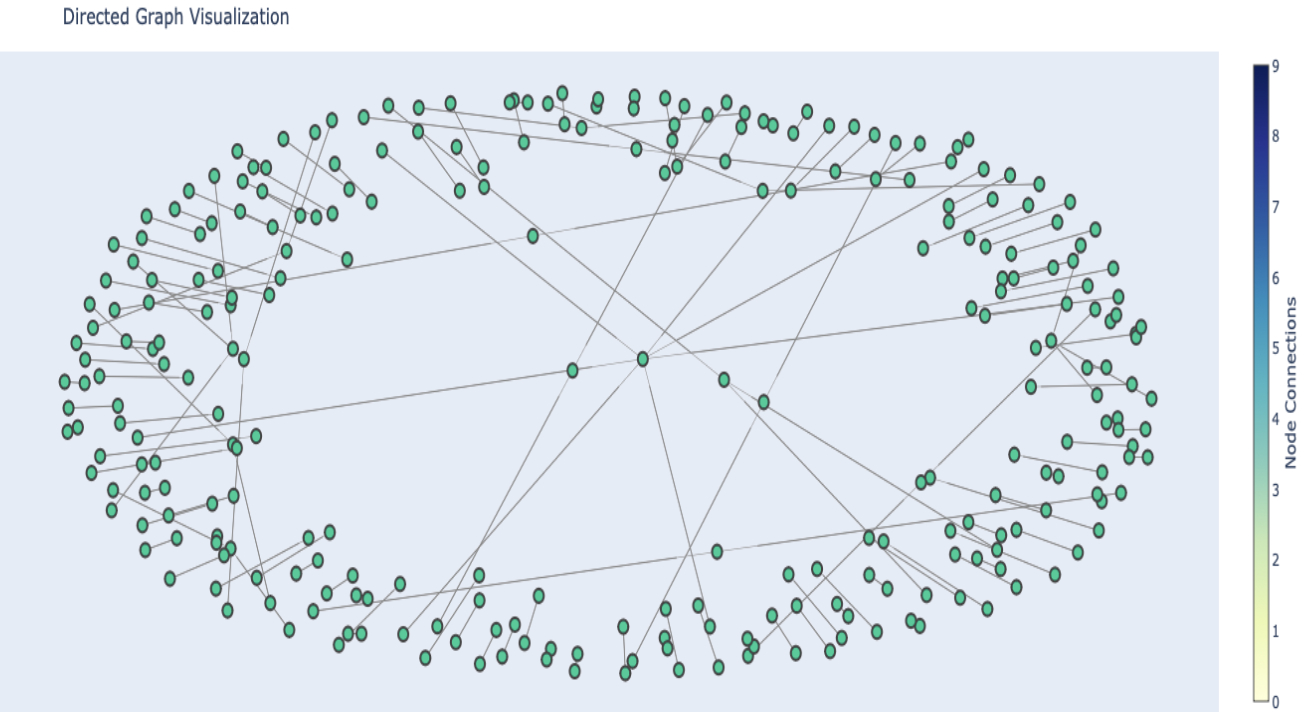}
    \caption{Graph network visualizer}
    \label{fig:graph_visualizer}
  \end{subfigure}
  \hfill
  \begin{subfigure}{0.48\textwidth}
    \centering
    \includegraphics[width=\linewidth]{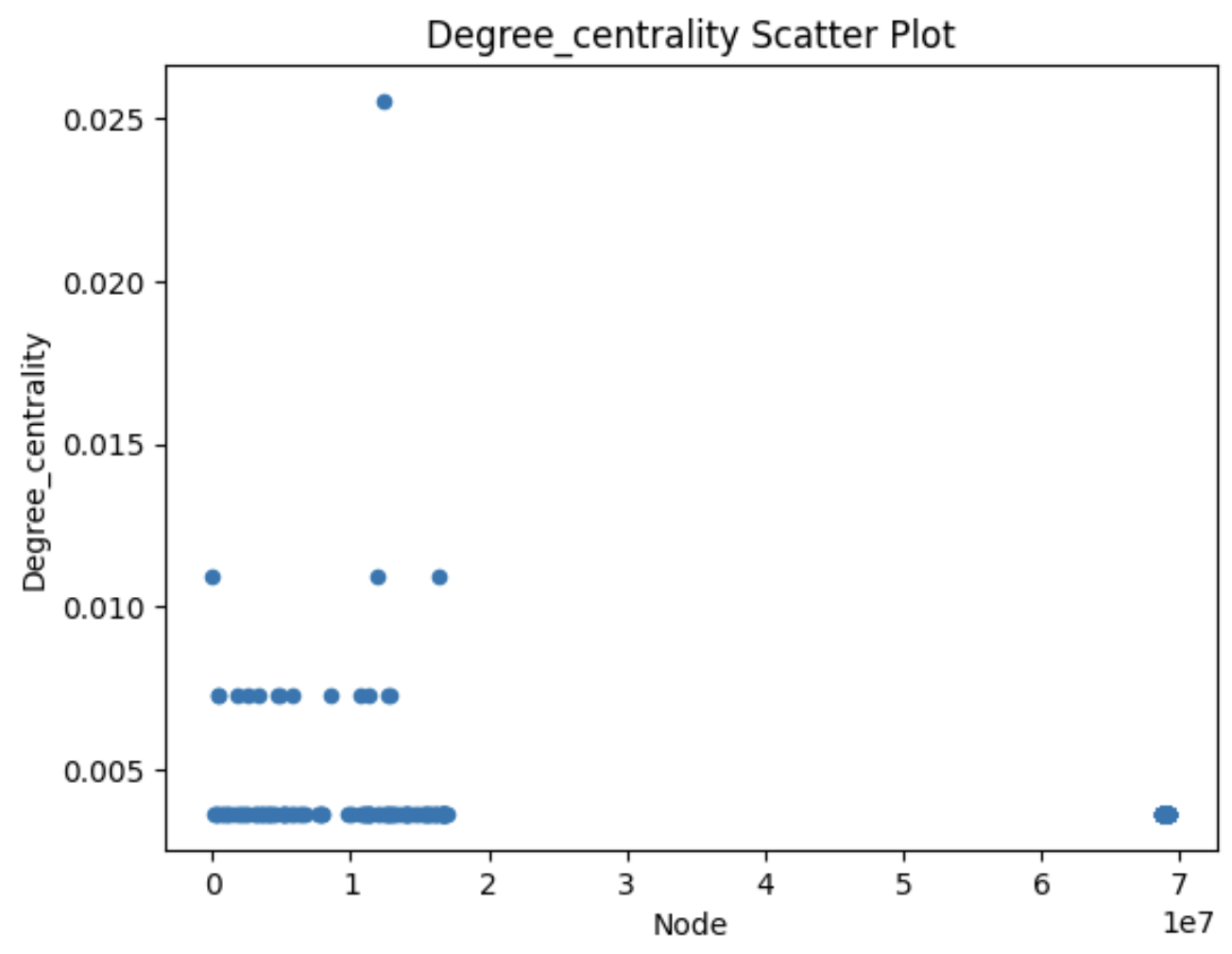}
    \caption{Graph network visualizer}
    \label{fig:graph_visualizer_2}
  \end{subfigure}
  \caption*{Figures 1: Graph network visualizer}
\end{figure}

\begin{figure}[h!]
  \begin{subfigure}{0.48\textwidth}
    \centering
    \includegraphics[width=\linewidth]{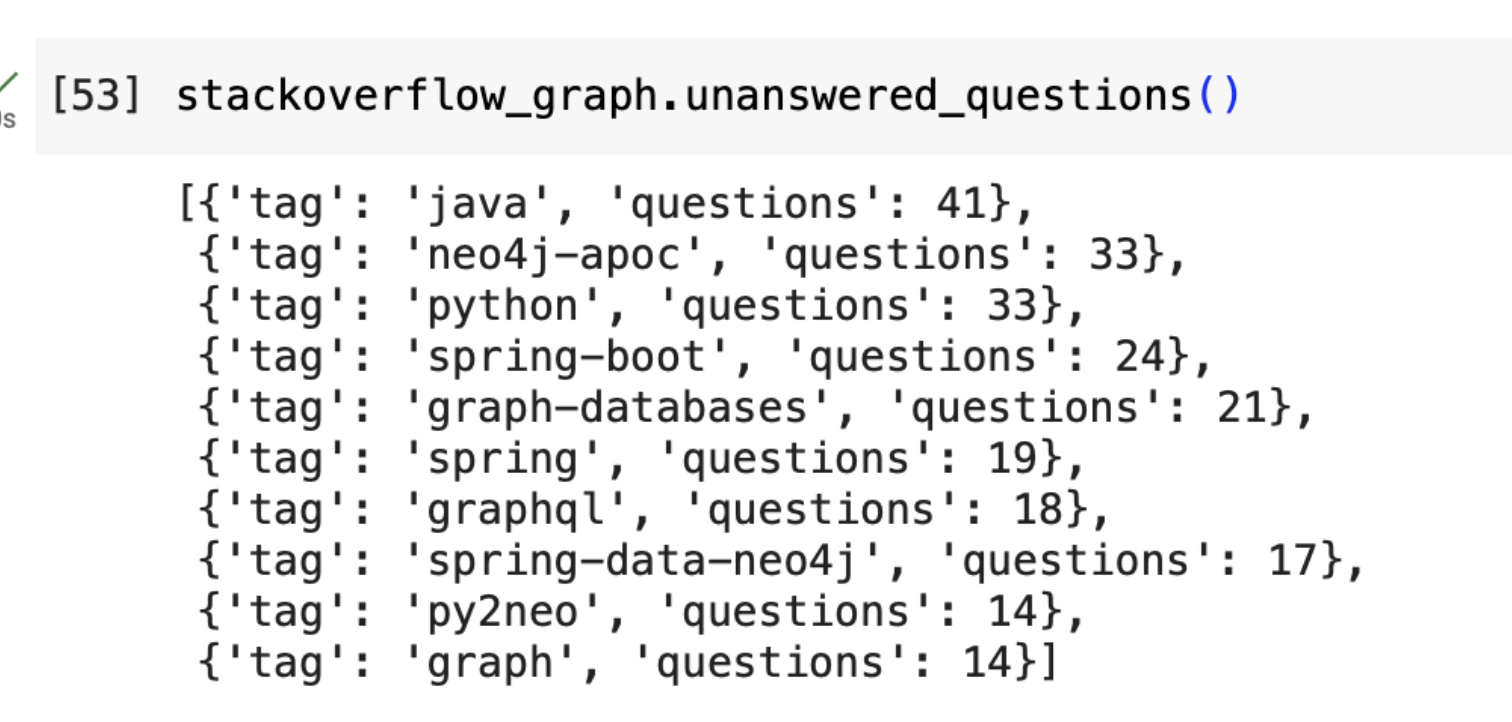}
    \caption{Unanswered Questions}
    \label{fig:unanswered_questions}
  \end{subfigure}
  \hfill
  \begin{subfigure}{0.48\textwidth}
    \centering
    \includegraphics[width=\linewidth]{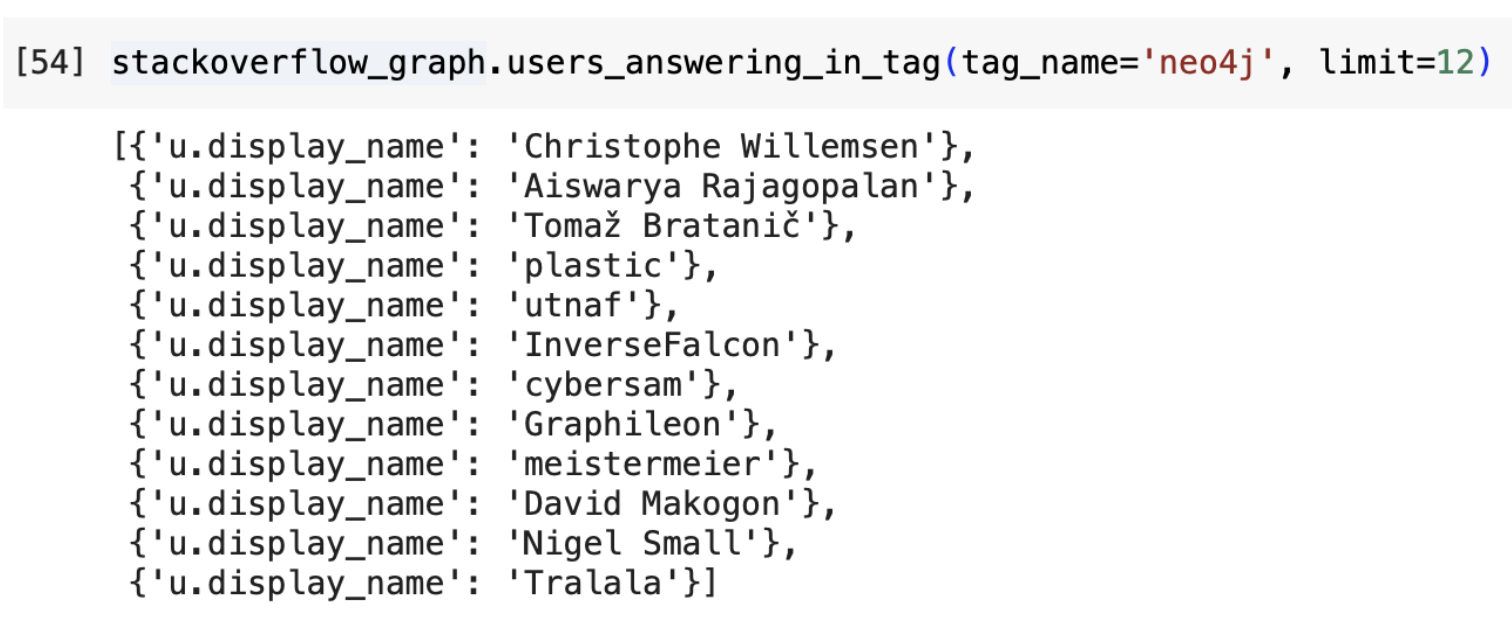}
    \caption{Users Answering in Tag}
    \label{fig:users_answering}
  \end{subfigure}
  \caption*{Figures 2a and 2b illustrate specific outcomes from the Neo4j analysis on Stack Overflow data}
    \begin{itemize}

\item (a) Shows the tags with the highest number of unanswered questions.
Item (b) Highlights users actively contributing answers within a specific tag.
Insights:

      \item Unanswered Questions: Identifying tags with many unanswered questions can help target areas where community engagement might be lacking, providing an opportunity to encourage more participation or focus efforts to resolve unanswered queries.
      \item Users Answering in Tag: Recognizing active contributors helps in identifying experts or highly engaged users within specific tags, which is crucial for community management and fostering a supportive environment.
    \end{itemize}
\end{figure}

\begin{figure}[h!]
  \centering
    \includegraphics[width=0.9\textwidth,height=0.9\textheight, keepaspectratio]{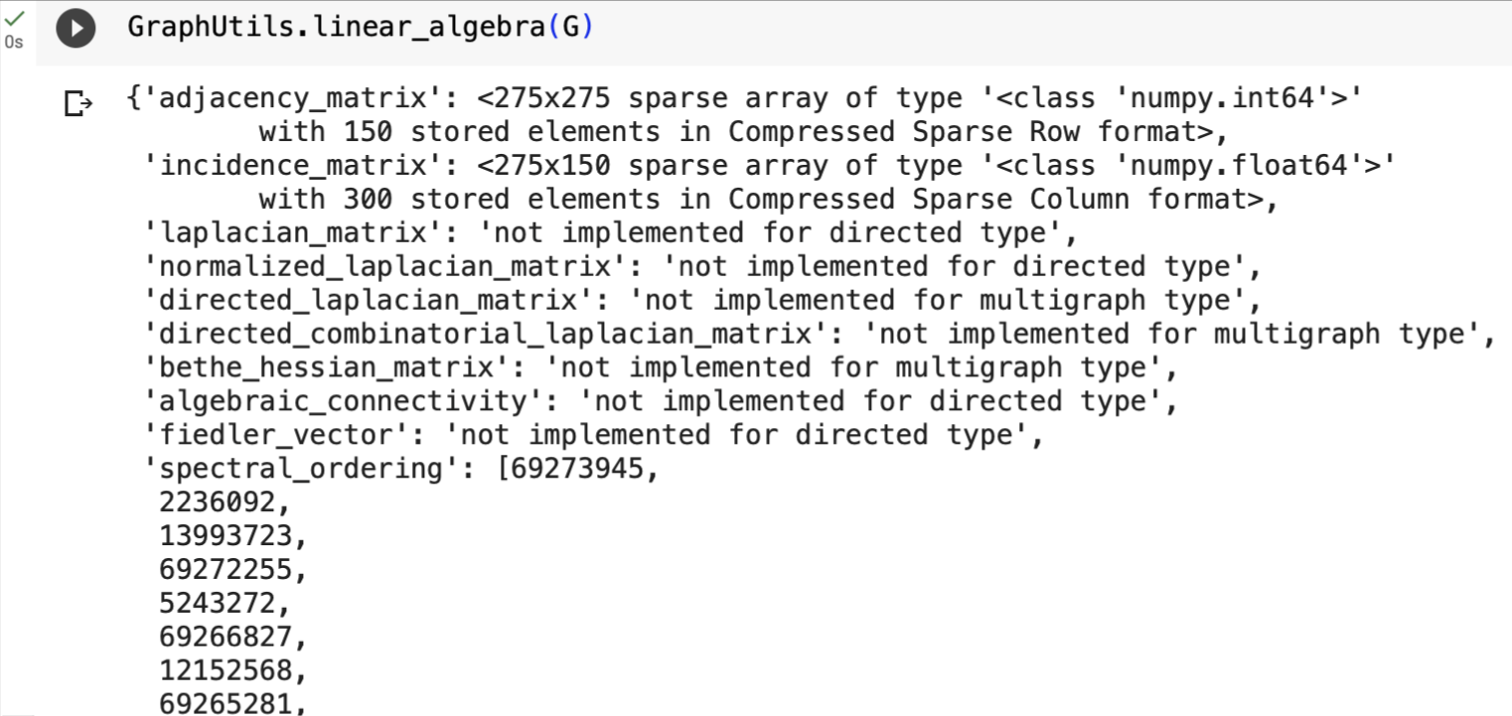}
  \caption{Application of linear algebra models to directed graphs: This figure demonstrates the importance of linear algebra models in preparing data for machine learning tasks. These models help in structuring and understanding complex network data, ensuring accurate data representation for advanced analytical techniques.}
  \label{fig:linear_algebra}
\end{figure}

\begin{figure}[h!]
  \centering
  \includegraphics[width=0.9\textwidth]{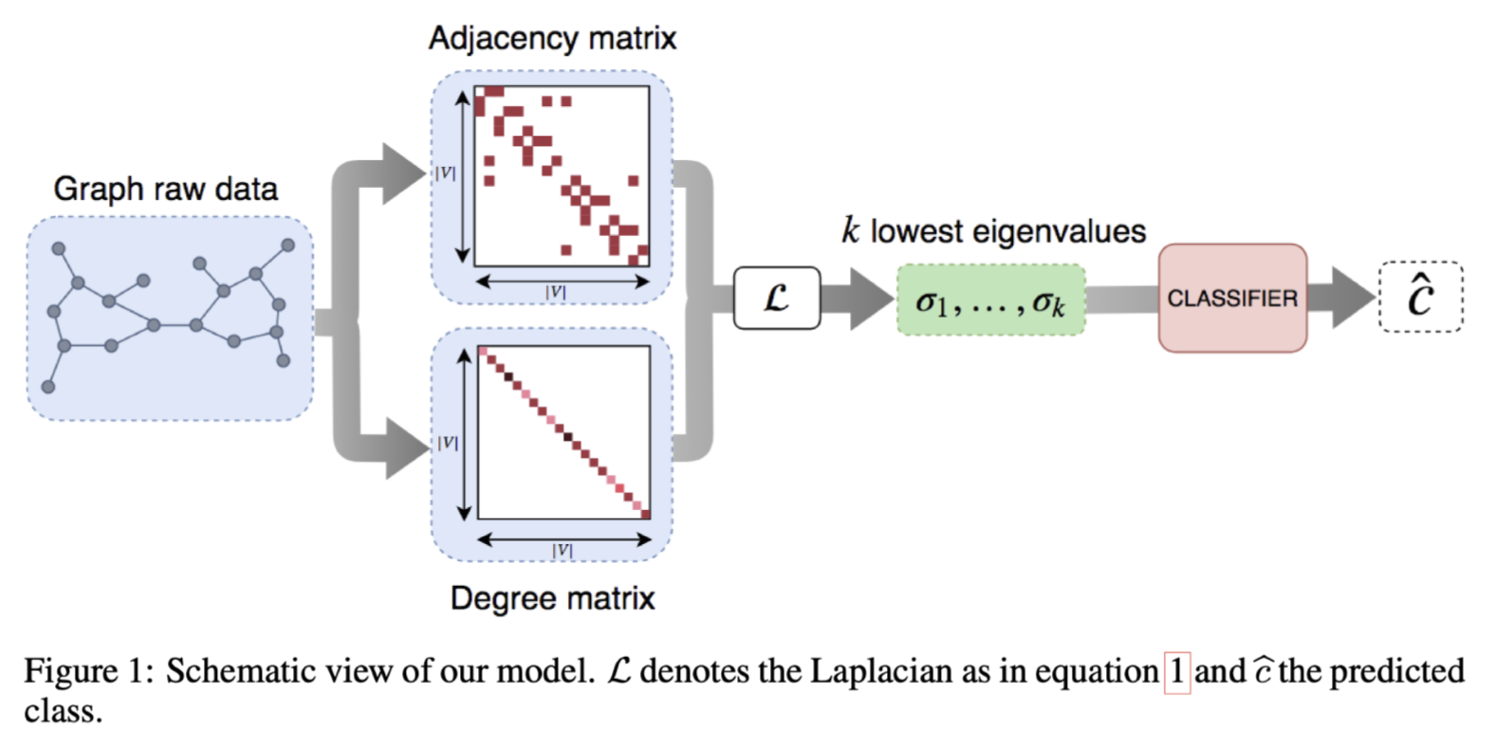}
  \caption{A Simple Baseline Algorithm for Graph Classification \cite{ref37}: The algorithm provides a foundational approach for more complex graph classification tasks, offering a benchmark against which other methods can be compared. It is a crucial step in developing more sophisticated models that can accurately classify and predict graph behaviors.}
  \label{fig:matrix_data}
\end{figure}

Figure \ref{fig:graph_visualizer} and \ref{fig:graph_visualizer_2} shows two visualizations of the network graph of Stack Overflow. These visualizations provide a graphical representation of the relationships between users and the structure of the network.

Insights:

The visualizations help identify key nodes and the overall connectivity of the network.
They illustrate the density of connections and can highlight central nodes that might play crucial roles in information dissemination and community interaction.

Additionally, Table \ref{table:node-degree-centrality} provides insights into the centrality of nodes within the network. The centrality measures obtained from this table reflect the relative importance of nodes in the network. After running the recommendation algorithm, we can expect similar centrality distributions, indicating that the detected items are closely related within the network.

\section{Graph Database Paradigm Shift and Community Detection}

The paradigm shift toward graph databases has imposed substantial pressure on well-established relational databases. In response to the challenges posed by big data, researchers have explored alternative database models, with graph databases emerging as promising candidates. This comparative analysis focuses on Neo4j, a leading graph database, within the context of community detection algorithms.

\subsection{A Simple Baseline Algorithm for Graph Classification}

Graph classification methods can schematically be divided into three categories: graph kernels, sequential methods, and embedding methods. In this section, we briefly present these different approaches, focusing on methods that only utilize the structure of the graph and no exogenous information, such as node features, to perform classification. This ensures a fair comparison of the algorithms' capacity to capture structural information \cite{ref37}.

The application of linear algebra models to directed graphs, as demonstrated in Figure \ref{fig:linear_algebra}, highlights the importance of structuring data for machine learning tasks. These models aid in understanding complex network data, ensuring accurate data representation for advanced analytical techniques.

Furthermore, Figure \ref{fig:matrix_data} illustrates the application of machine learning or deep learning techniques in graph classification. While our proposed simple baseline algorithm is based on spectral decomposition, this figure showcases the broader landscape of methods available for graph classification, emphasizing the diversity of approaches in the field.

\subsection{Community Detection in Networks}

Networks play a fundamental role in representing complex data structures with entities and intricate relationships. Graphs, as a powerful representation for networks, enable visual inspection and the application of graph algorithms to extract valuable insights. A critical property of many real-world networks is their inherent heterogeneity, known as "community structure."

Community detection, or graph clustering, is the process of identifying hidden patterns and distinct communities within a graph. This process has practical implications, such as resource management in computer networks, customer segmentation based on buying behavior, and identifying influential clusters in social media networks.

\subsection{Challenges of Big Data in Community Detection}

In the era of Big Data, networks often represent massive datasets with millions or billions of vertices, amplifying the complexity of community detection. This demands novel approaches to outperform traditional sequential and centralized community detection algorithms.

\subsection{Neo4j: A Graph Database for Community Detection}

This comparative analysis focuses on Neo4j, a prominent graph database, assessing its potential for community detection on large graphs. Neo4j's unique advantage lies in its ability to execute graph-related algorithms directly on stored data without the need for data transformation.

\subsection{Experimental Setup and Methodology}

To evaluate Neo4j's capabilities for community detection, a set of experiments was designed. The Label Propagation Algorithm (LPA) was chosen for its low computational complexity and compatibility with both Neo4j and Apache Spark. The experiments utilized datasets from the Stanford Large Network Dataset Collection (SNAP), offering varying sizes and complexities. As shown in Figure \ref{fig:linear_algebra}, the application of linear algebra models to directed graphs demonstrates their importance in preparing data for machine learning tasks, ensuring accurate data representation for advanced analytical techniques.

\subsection{Results and Discussion}

The comparative analysis using Neo4j provided noteworthy findings. Overall, Neo4j's engine demonstrated superior performance compared to the Apache Spark cluster for most selected datasets. Its efficiency in processing and analyzing community structures within graphs aligned with the research objectives.

\subsubsection{Results Obtained from Neo4j Code}

\begin{itemize}
    \item \textbf{Top Tags:} Neo4j identified the most prevalent tags in the Q{\&}A platform, with "neo4j" and "cypher" among the top tags based on the number of associated questions. This analysis is further illustrated in Figures \ref{fig:graph_visualizer} and \ref{fig:graph_visualizer_2}.
    
    \item \textbf{Unanswered Questions:} Neo4j efficiently identified a list of tags with the highest number of unanswered questions, providing insights into areas where the community may need more engagement. Figure \ref{fig:unanswered_questions} illustrates specific outcomes from this analysis.
    
    \item \textbf{Users Answering in Tag:} Neo4j pinpointed users actively contributing answers within a specific tag, facilitating the recognition of community experts and contributors. This is depicted in Figure \ref{fig:users_answering}.
\end{itemize}

\section*{Analysis of Neo4j and Apache Spark}

However, it's essential to note that there exists a tipping point where the size of the processing dataset surpasses a certain threshold. In such cases, Apache Spark exhibited better performance in handling larger datasets. This observation suggests that while Neo4j excels in many aspects of community detection, scalability becomes a concern when dealing with extremely large graphs.

Another significant advantage identified during the experiments was the user-friendliness of Neo4j. The installation process for Neo4j was notably easier and more straightforward compared to Apache Spark, further enhancing its appeal for researchers and practitioners seeking efficient solutions for graph-related tasks.

\subsection*{Conclusion and Future Directions}

In conclusion, Neo4j emerges as a robust environment for storing and analyzing graph data, making it an ideal choice for community detection tasks, particularly with moderately sized datasets. Its ability to execute graph-related algorithms directly on stored data streamlines the analysis process and enhances its usability. However, it's essential to recognize the limitations of Neo4j, particularly concerning scalability when dealing with massive graphs. In such cases, distributed environments like Apache Spark become necessary to achieve optimal performance.

As a direction for future research, exploring alternative graph algorithms and evaluating the performance of Neo4j in conjunction with distributed solutions, such as the Neo4j Connector for Apache Spark, could offer a comprehensive understanding of how to harness the strengths of both systems. Overall, this comparative analysis highlights Neo4j's potential as a valuable tool for community detection in large graphs, providing valuable insights into the complex structures inherent in real-world networks.

\section*{Linear Algebra and Machine Learning}

We will explore the utilization of linear algebra data for various analytical purposes. This dataset comprises critical matrices and associated metrics, including the adjacency matrix, incidence matrix, Laplacian matrix, normalized Laplacian matrix, directed Laplacian matrix, directed combinatorial Laplacian matrix, Bethe Hessian matrix, algebraic connectivity, Fiedler vector, and spectral ordering. These matrices and metrics play a pivotal role in understanding and characterizing complex networks and graph structures.

\subsection*{Adjacency Matrix}

This matrix represents the connections between nodes in a graph, with non-zero entries indicating the presence of an edge between nodes. It is fundamental for various graph-based algorithms and analyses.

\subsection*{Incidence Matrix}

The incidence matrix is used to represent the relationships between nodes and edges in a graph. It is particularly useful for analyzing graphs in terms of node-edge relationships.

\subsection*{Laplacian Matrix}

Although not implemented for directed graphs, the Laplacian matrix is a crucial tool for studying graph connectivity and spectral properties.

\subsection*{Normalized Laplacian Matrix}

Like the Laplacian matrix, the normalized Laplacian matrix provides insights into the spectral properties of a graph, with eigenvalues falling in the range [0, 2]. It is often employed for machine learning tasks without the need for extensive preprocessing.

\subsection*{Directed Laplacian Matrix}

Although not implemented for multigraphs, the directed Laplacian matrix is essential for understanding directed graphs and their spectral characteristics.

\subsection*{Directed Combinatorial Laplacian Matrix}

Similar to the directed Laplacian matrix, this matrix is also significant for directed graph analysis.

\subsection*{Bethe Hessian Matrix}

This matrix is useful for understanding certain graph structures and properties, although it is not implemented for multigraphs.

\subsection*{Algebraic Connectivity}

This metric provides insights into the connectedness of a graph and is a crucial indicator of its robustness.

\subsection*{Fiedler Vector}

The Fiedler vector is used in spectral graph theory and graph partitioning, aiding in the analysis of graph structures.

\subsection*{Spectral Ordering}

Spectral ordering, as shown in the provided data, can help in visualizing and understanding the organization of nodes within a graph.

This comprehensive dataset allows us to perform advanced analyses on various types of graphs and networks, ranging from social networks to biological networks and beyond. The inclusion of spectral properties, connectivity metrics, and ordering information enhances our ability to extract meaningful insights from complex data.

\subsection*{Reference}

The paper "A Simple Baseline Algorithm for Graph Classification" by Nathan de Lara and Edouard Pineau, which provides valuable insights into the spectral properties of normalized Laplacian eigenvalues and their applications in graph classification. The paper discusses how these eigenvalues can be leveraged for downstream tasks, such as classification, and highlights their relevance in characterizing graph structures.

Furthermore, the paper connects the eigenvalues of the Laplacian to energy levels and frequency decomposition in the context of stable configurations and signal processing on graphs. It also hints at the potential connection between spectral decomposition and graph isomorphism, which remains an open research problem.

In conclusion, the research presented in our thesis demonstrates the significance of normalized Laplacian eigenvalues in graph classification. We have shown that this feature can be easily extracted and integrated into existing graph representation models, offering the potential for improved performance in a wide range of applications.

\section*{Concluding Remarks and Future Directions}

The comparative analysis reveals that while traditional systems offer accuracy and variety in terms of algorithmic selection, neuromorphic systems have the potential to revolutionize the scalability and energy efficiency aspects. Future work should explore:

\begin{itemize}
  \item \textbf{Hybrid Systems:} Combining the robustness of traditional algorithms with the speed and energy efficiency of neuromorphic computing.
  \item \textbf{Advanced Algorithms:} Implementing complex graph algorithms on neuromorphic systems.
  \item \textbf{Personalization and Machine Learning:} Integration of machine learning techniques for more personalized, actionable insights.
  \item \textbf{Optimization Techniques:} Application of advanced optimization methods to enhance the efficiency of traditional algorithms.
\end{itemize}

\section*{Comparative Evaluation Extension}

\begin{itemize}
  \item \textbf{Real-Time Performance:} In the area of real-time performance, neuromorphic systems outshine the Graph Analysis Engine. Their inherent parallelism allows them to handle live-streaming data effectively, making them ideal for applications such as real-time anomaly detection in network security.
  \item \textbf{Algorithmic Limitations:} While neuromorphic systems offer energy efficiency and scalability, they are not without limitations. As they are in the exploratory phase of their development, their ability to accurately implement complex graph algorithms is still an open question. Traditional systems still retain an edge in terms of the maturity and robustness of their algorithms.
\end{itemize}

\section*{Conclusions}

The existing literature on the use of social network analysis techniques and graph algorithms to study the Stack Overflow community provides a good understanding of the structure of the community, the flow of information, and the importance of certain users and questions. However, there is a gap in the research on the content of the questions and answers and how this content changes over time, and a need for further research using graph network flow analysis to study the flow of information within the community. The existing literature provides a consistent view of the Stack Overflow community as a highly clustered network with a few central users who play a key role in the flow of information.

\end{document}